%% file: paper.tex
\documentclass[conference]{IEEEtran}

\usepackage{amsmath}
\usepackage{graphicx}
\usepackage{epstopdf}
\usepackage{multirow}

\graphicspath{{images/}}

\usepackage[justification=centering, labelfont=bf]{caption}
\usepackage{subcaption}
\usepackage{listings}
\usepackage{verbatim}
\usepackage{url}
\usepackage{footnote}
\usepackage{array}
\usepackage{hhline}
\usepackage{tikz}
\usepackage{dingbat}

\usepackage{algorithm}
\usepackage[noend]{algpseudocode}

\let\origthelstnumber\thelstnumber
\makeatletter
\newcommand*\Suppressnumber{%
  \lst@AddToHook{OnNewLine}{%
    \let\thelstnumber\relax%
     \advance\c@lstnumber-\@ne\relax%
    }%
}

\newcommand*\Reactivatenumber[1]{%
  \setcounter{lstnumber}{\numexpr#1-1\relax}
  \lst@AddToHook{OnNewLine}{%
   \let\thelstnumber\origthelstnumber%
   \refstepcounter{lstnumber}
  }%
}

\def\setspacing#1{\renewcommand{\baselinestretch}{#1}\small\normalsize}
\setspacing{0.99}
\graphicspath{{./figures/}}

\begin{document}

\title{A Compilation Flow for the Generation of \\ CNN Inference Accelerators on FPGAs}

\author{\IEEEauthorblockN{Seung-Hun Chung and Tarek S.\ Abdelrahman}
\IEEEauthorblockA{The Edward S. Rogers Sr.\ Dept.\ of Electrical and Computer Engineering\\
University of Toronto\\
Toronto, Ontario Canada M5S~3G4\\
Email: sh.chung@mail.utoronto.ca, tsa@ece.utoronto.ca}}

\maketitle

\input{abstract}

\input{introduction}

\input{background}

\input{flow}

\input{optimizations}

\input{evaluation}

\input{related}

\input{conc}

\bibliographystyle{IEEEtran}
\bibliography{paperbib}

\end{document}

%% file: abstract.tex
\begin{abstract}
We present a compilation flow for the 
generation of CNN inference accelerators on FPGAs. The flow 
translates a frozen model into OpenCL kernels with the TVM compiler 
and uses the Intel OpenCL SDK to compile to an FPGA bitstream. 
We improve the quality of the generated hardware with optimizations
applied to the base OpenCL kernels generated by TVM. These
optimizations increase parallelism, reduce memory access latency, 
increase concurrency and save on-chip resources. 
We automate these optimizations in TVM and evaluate them
by generating accelerators for LeNet-5, MobileNetV1 and ResNet-34
on an Intel Stratix~10SX. We show that the optimizations improve 
the performance of the generated accelerators by up to 
846$\times$ over the base accelerators. The performance of the optimized
accelerators is up to 4.57$\times$ better than TensorFlow on CPU, 
3.83$\times$ better than single-threaded TVM and is only 0.34$\times$ 
compared to TVM with 56 threads. Our optimized kernels also outperform
ones generated by a similar approach (that also uses high-level synthesis) 
while providing more functionality and flexibility. However, it 
underperforms an approach that utilizes hand-optimized designs. Thus, 
we view our approach as useful in pre-production environments that 
benefit from increased performance and fast prototyping, realizing
the benefits of FPGAs without hardware design expertise.
\end{abstract}

%% file: introduction.tex
\section{Introduction}

Field Programmable Gate Arrays (FPGAs) offer unique opportunities 
to realize accelerators for Convolutional Neural Network (CNN)
inference. They enable customized network-specific accelerators
that are more resource efficient than generic accelerators,
and they facilitate rapid prototyping of accelerators for new
and evolving networks. However, they come with their unique 
challenges. FPGAs are programmed using a hardware design model, 
often at the register transfer level (RTL), which is foreign to 
most machine learning experts and software developers. Thus,
there is room for tools that can ease the deployment of
CNN accelerators to FPGAs.

We propose, implement and evaluate an approach to the automatic 
generation of FPGA CNN inference accelerators. The approach utilizes 
the direct mapping of layers to FPGA hardware (i.e., direct dataflow 
synthesis~\cite{xilinx-finn,venieris2016fccm}). The frozen graph 
representation of a CNN is compiled into OpenCL kernels that 
perform the computations of network layers. 
An OpenCL-to-RTL compiler translates these kernels into RTL, which 
is then synthesized into FPGA hardware. This approach contrasts to 
the one adopted by most current state-of-the-art accelerators that
use either a spatial array of processing elements (PEs) that is
time multiplexed over the layers of the network, or a set of
pre-designed and optimized hardware templates.

Our approach facilitates rapid generation of CNN accelerators
using a compiler-based flow that does not expose users to hardware 
design. It also generates network-specific accelerators that 
can utilize network and even layer-specific bit precisions, 
resulting in more efficient hardware and potentially better 
performance than generic PE-based ones~\cite{Guo2019}. 
Further, it alleviates the need for hand-designed templates 
that require updates for new network operations, or new
layer precisions. Finally, its use of OpenCL promotes 
portability across FPGA devices.

However, compiler-generated OpenCL kernels perform poorly due to
lack of parallelism in generated hardware and/or poor utilization 
of memory bandwidth. Worse, the kernels may not synthesize at all for 
larger networks, where the design exceeds the target FPGA resources. 
Thus, we utilize a set of optimizations to improve performance and
conserve resources. These optimizations include loop tiling, strip 
mining, unrolling, fusion, parameterized kernels, auto-run kernels, 
and concurrent kernel execution.

We prototype our approach and optimizations using the TVM
compiler~\cite{ApacheTVM} and the Intel OpenCL 
SDK~\cite{IntelProgrammingManual}. We show the 
viability of our approach by generating accelerators 
for three CNNs: LeNet-5, MobileNetV1, and ResNet-34 on an Intel 
Stratix~10SX. We experimentally measure the impact of the 
optimizations and show that they improve performance over 
the base OpenCL kernels accelerators by factors of 10$\times$, 
184$\times$ and 846$\times$ for the three networks respectively. 
The optimized accelerators show improvement over TensorFlow on
a CPU by up to 4.57$\times$, over single-threaded TVM by up to
3.83$\times$ and a slowdown of about 0.34$\times$ compared to
TVM with 56 threads. The performance of the accelerators is 
better than that of a GTX~1060 for the smaller LeNet-5 but is 
worse for the larger networks.
Further, our optimized kernels outperform by about 1.4$\times$ 
ones generated by a similar approach that also uses high-level 
synthesis, but is limited to 3$\times$3 convolutions~\cite{7929549}.
However, it underperforms an approach that utilizes hand-optimized 
templates~\cite{10.1145/3186332} by 9.22$\times$.

Our results lead us to conclude that: (1) our optimizations are
effective in improving the performance of the generated hardware
over the base OpenCL hardware, (2) the performance of the 
generated accelerators is competitive with that of a CPU and
a GPU (for smaller networks), and (3) as expected, 
our accelerators underperform hand-optimized designs. Thus,
we view our approach as best suited for pre-production environments 
that benefit from increased performance and fast prototyping,
realizing the benefits of customizable hardware without 
hardware design expertise.

The remainder of this paper is organized as follows. We introduce 
the two tools critical to our work in Section~\ref{sec:background}.
Our compilation flow is presented in Section~\ref{sec:flow} and 
the optimizations we use are described in Section~\ref{sec:optimizations}.
The evaluation is given in Section~\ref{sec:evaluation}. 
Finally, related work is reviewed in Section~\ref{sec:related}
and concluding remarks are presented in Section~\ref{sec:conc}.

%% file: background.tex
\section{Background}
\label{sec:background}

\subsection{TVM}
\label{sec:tvm}

The Apache Tensor Virtual Machine (TVM)~\cite{ApacheTVM} is an open-source 
compiler framework that compiles and optimizes deep learning 
models from high-level machine learning frameworks and deploys them on a 
variety of target lower-level languages (e.g., LLVM IR, CUDA, OpenCL).

TVM imports models from deep learning frameworks into the top-level 
functional intermediate representation called {\em Relay} 
IR~\cite{DBLP:journals/corr/abs-1810-00952}. 
It then applies rules-based transformations such as operator fusion, dead code elimination, 
and layout changes. Relay operations are then lowered to {\em tensor expressions},
which is a domain-specific language for kernel construction. These 
expressions are translated into code using \textit{compute functions} that 
implement common tensor expression operators (e.g., 2D convolution and 
fully-connected layers).

Optimizations are made to the compute functions through a \textit{schedule}.
Scheduling primitives include loop unrolling, fusion, tiling, 
and vectorization. TVM maintains a library of compute functions that 
implement common tensor expression operators as well as schedules for 
common platforms.

\subsection{Intel OpenCL FPGA Compiler}

The Intel AOC compiler~\cite{IntelProgrammingManual} is used to compile 
OpenCL kernel programs into a bitstream for FPGAs. This
bitstream implements both lower-level utility (``shell'') logic and user kernel logic. 
A host program manages the creation and movement of buffers and
tasks for execution.

AOC implements constant, private, and local memory in registers or 
in embedded memory blocks (BRAM). Global memory is implemented in external 
memory (DDR4). Accessing data implemented in BRAMs and external 
memory requires the generation of load-store units (LSUs). These LSUs are 
implemented in logic elements (and BRAMs if caches are inferred). 

There are several types of LSUs inferred by the compiler,
depending on the type of access and the size of the arrays in memory. 
They include: coalesced, burst-coalesced, prefetching and streaming 
pipelined~\cite{IntelBPGuide}. These LSUs vary in performance and resource 
usage, with coalesced and burst-coalesced being the most efficient. They 
are inferred when arrays are accessed are aligned and consecutive.

%% file: flow.tex
\section{Compilation Flow}
\label{sec:flow}

\begin{figure*}[ht!]
  \centering
  \includegraphics[width=0.7\linewidth]{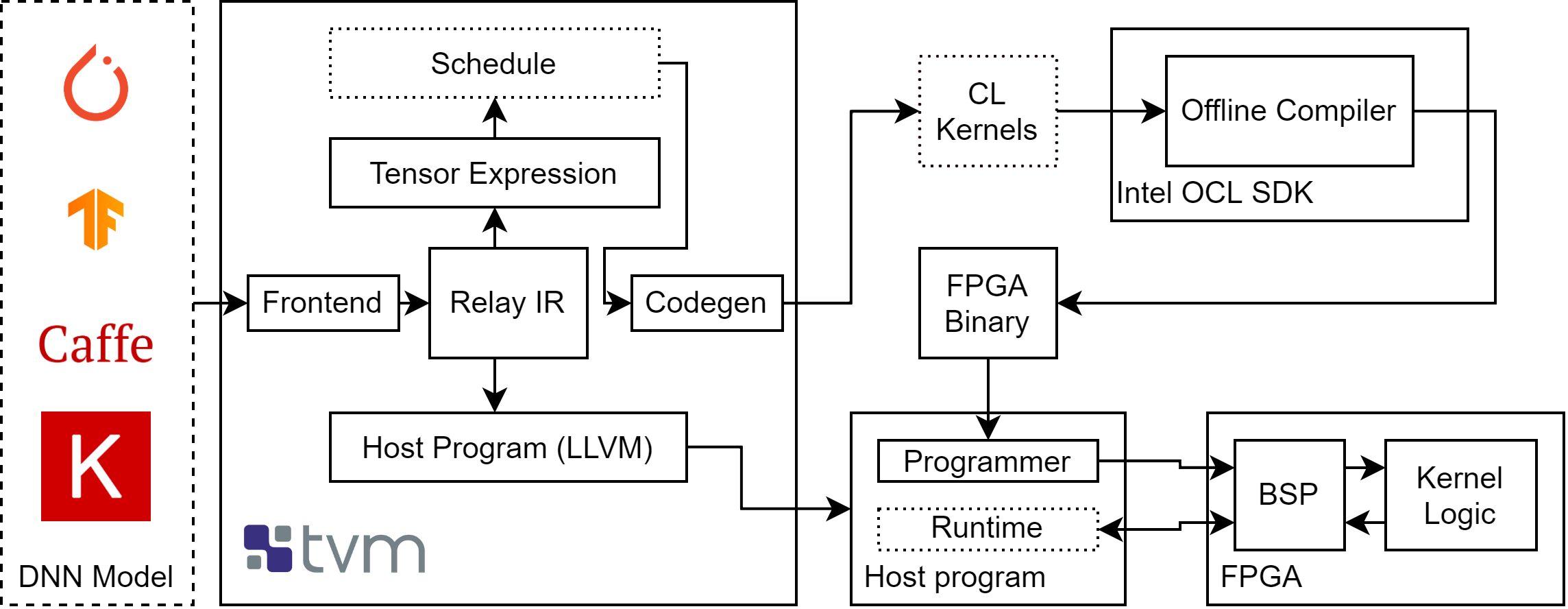}
  \caption{The compilation flow}
  \label{fig:deployment:flow}
\end{figure*}

A high level overview of our compilation flow is shown in 
Fig.~\ref{fig:deployment:flow}. It begins with a graph 
representation of a CNN that is trained in any of the high-level 
ML frameworks that TVM supports (e.g., PyTorch, TensorFlow). 
We apply our optimizations to the schedules that
target the AOC compiler. Specifically, we modify the components
of TVM that are encased in dotted lines in 
Figure~\ref{fig:deployment:flow}), which include the schedule, 
generated kernels, and the host runtime.  

Our flow supports two modes of execution: \textit{pipelined} and 
\textit{folded}. In pipelined execution, output feature 
maps are transferred between kernels using OpenCL channels. A kernel 
is generated per layer, and all kernels are active concurrently. To 
pipeline inference, all activations of a layer must be stored in a 
kernel's local memory. Given the scarcity of on-chip storage on FPGAs, 
this limits deployment to smaller networks. 

However, it is common for CNNs to have a ``workhorse'' operation that is repeated. 
For example, $1\times 1$ convolutions constitute 94.9\% of multiply-adds 
in MobileNetV1~\cite{howard2017mobilenets}. This makes it possible to 
re-use parameterized (dynamic input/output tensor shapes) kernels for 
multiple layers in the network, consuming less FPGA resources compared 
to a fully pipelined design. We refer to this as {\em folded} execution, and 
use this to deploy larger networks such as MobileNetV1 and ResNet-34. 
Although it is theoretically possible to pipeline a network that is also 
folded, it is unlikely that there would be sufficient on-chip memory to 
implement both.

Our flow has several benefits:

\begin{itemize}
\item It enables the automatic generation of CNN accelerators, isolating
user from hardware design and reducing development time while being 
flexible to implement different operations. 

\item It builds upon industry-standard tools; the use of Intel's OpenCL 
SDK for FPGAs promotes portability across Intel FPGA platforms, including 
upcoming ones. 

\item It uses an open-source ML compiler to leverage community support
for cutting-edge frameworks and computational operations. Adding support 
for new operators will only require an optimization of its schedule in 
this flow. In contrast, with flows such as DnnWeaver~\cite{DNNWeaver}, 
fpgaConvNet~\cite{venieris2016fccm}, 
and MALOC~\cite{MALOC}, deploying other types of networks may require some 
overhaul of the accelerator architecture.
\end{itemize}

\noindent
However, there are challenges and limitations to this approach:

\begin{itemize}
\item Representing layers as OpenCL kernels limits the kernels to 
synchronize at the granularity of the entirety of input/output feature 
maps. Further, like other dataflow accelerators, on-chip resource 
constraints must be considered when mapping network execution to kernels.

\item Circuits generated by TVM/AOC will likely not be competitive with 
designs generated from RTL/HLS templates made by hand, in terms of clock 
frequency and hardware utilization.

\item Designs generated from naive schedules perform poorly, underutilize 
hardware, and inefficiently use global memory, requiring the optimizations 
that we perform. We propose and implement a set of optimizations to 
alleviate this limitations, and they are the focus of this paper.
\end{itemize}

We believe the combination of an easy-to-use ML interface and the 
flexibility gained from generating kernels automatically from a tensor 
language may outweigh the downsides of the approach after the 
optimizations.

%% file: optimizations.tex
\section{Optimizations}
\label{sec:optimizations}

Kernels generated from the default schedule for the AOCL target
in TVM perform poorly or fail to synthesize due to several reasons:
\begin{enumerate}
\item Activations and normalizations are computed in a loop
adjacent to convolutions, which gives rise to read-after-write dependences. 
These dependences prevent loop pipelining.

\item The generated kernels do not produce parallel hardware
since loops are not automatically unrolled. This 
results in only few arithmetic operations performed per cycle.

\item External memory bandwidth is underutilized. Further, 
the generated kernels use global memory for all data including 
accumulations. Data can be moved locally to utilize higher on-chip memory bandwidth.

\item A one-to-one layer-to-kernel mapping can easily exhaust 
resources due to excessive logic usage for LSUs, preventing 
a design to synthesize for larger networks.
\end{enumerate}

Thus, we apply a number of optimizations to achieve a higher level 
of parallelism, increase global memory utilization, and
reduce resource usage. We group these optimizations into two 
categories: \textit{Kernel Optimizations} (transformations 
made to an OpenCL kernel code) and \textit{Host Optimizations} 
(modifications made to the OpenCL host program).

\subsection{Loop Unrolling}
\label{sec:unroll}

Loop unrolling replicates a loop body in a kernel by a specified 
unroll factor, $f$. It has three effects. First, it replicates 
compute logic (DSPs) by $f$ in the loop body, which increases parallelism. 
Second, it widens the LSU access width by $f$ when global memory 
access patterns are consecutive, which increases efficiency of
memory operations. Finally, for loads/stores to local memory, it 
replicates or banks the BRAM to create more read/write ports and 
allow concurrent read/writes.

We only fully unroll loops since partial unrolling may limit
performance gains~\cite{IntelProgrammingManual}. In the case of a 
loop that has too many iterations to fully unroll, we strip mine
and fully unroll the resulting inner loop.

The choice of the unroll factor is crucial. A large factor 
increases parallelism and memory utilization, and therefore performance. 
However, excessive unrolling exhausts on-chip resources. 
For example, replicating non-consecutive accesses leads to LSU 
replication, which incurs a significant cost in logic and BRAM. 
Similarly, excessive replication of BRAM adds logic for memory 
arbitration and stalls accesses. Finally, excessive unrolling may 
result in a circuit that exceeds memory bandwidth, having LSUs 
compete for memory bandwidth and stall, greatly 
degrading performance~\cite{DBLP:journals/corr/abs-1810-09773}. 

\subsection{Loop Strip Mining/Tiling}
\label{sec:strip}

Loop strip mining/tiling controls the granularity of a loop by 
partitioning a loop into smaller segments (or strips)~\cite{Bacon}. 
On their own these optimizations are not desirable for FPGA kernels 
since the extra loop(s) they create causes extra loop control logic 
to be generated. Further, if the loop count is not evenly divisible 
by the tiling factor, code to complete the remaining iterations 
must be placed after the strip mined/tiled loops. 
Thus, we apply these transformations with the intent that resulting 
inner loops are to be fully unrolled, achieving the same effect as 
partial loop unrolling.

\subsection{Loop Fusion}
\label{sec:loopfusion}

Loop fusion merges two adjacent loops into a single loop. It has the 
effect of reducing loop control structures, and thus resource usage.
It can also decrease storage requirements.  For example, many kernels
have a reduction loop that stores the result in a temporary array, 
followed by a loop that performs the activation function using this 
array. By fusing the two loops, it becomes unnecessary to use the 
array; rather only a local register is used. This decreases global 
memory contention by removing the LSUs for accessing the temporary 
array and frees memory bandwidth. 

\subsection{Cached Writes}
\label{sec:cachedwrite}

When possible writes should be cached in a scratchpad to avoid the
generation of LSUs. This is especially important for the common
pattern of accumulations, where the previous sum is loaded before
adding a new result. Such accumulations can be optimized by replacing
the memory variable by a local variable that resides in a local
register.
Cached writes are implemented in TVM by adding an extra stage to the
computation. The original computation is performed using local
memory and the added stage copies data from local memory to global
memory.

\subsection{Channelization}
\label{sec:channel}

OpenCL channels allow direct kernel-to-kernel communication without 
the use of global memory and without host control. Thus, channels
place direct paths between communicating kernels and implement
these paths using registers, avoiding the use of global memory. 
Channels lead to a marked reduction in the number of LSUs since 
kernels communicate on chip. This not only reduces the resource 
usage but also reduces contention for global memory. 

Channel reads stall when attempting to read from an empty channel
or write to a full channel. Thus, unequal consumer/producer rates 
cause stalls in the kernel that degrades performance. This can be 
alleviated by using buffered channels, where a FIFO queue with a 
user-specified depth is added to the channel.

Channels requires only modifications to the host program.
Nonetheless, their use may require changes to kernel code. For example,
if a kernel needs to re-use data that it consumed from a channel, 
it needs to store channel reads into local memory (registers or 
BRAM), where it can be re-accessed. 

\subsection{Autorun Kernels}
\label{sec:autorun}

Kernels that have no arguments (i.e., no accesses to global 
memory) can be declared \textit{autorun}. Such kernels execute 
independent of the host, and are equivalent to wrapping a kernel 
with a \texttt{while(true)}. This execution mode decreases the 
overhead of command queue control from software and is most 
beneficial when kernel execution times are small compared to
kernel launch overhead. Combined with the use of channels, autorun
kernels facilitate the streaming on our accelerators.

\subsection{Concurrent Execution}

A single command queue for queuing tasks to a device serializes 
kernel execution. For multiple kernels to run in parallel, multiple 
command queues are required, one for each kernel. Concurrently 
executing kernels increase hardware utilization because multiple 
kernels are active at the same time. However, if dependences 
exist between kernels, they must be synchronized using channels.

\subsection{Parameterized Kernels}

TVM generates an OpenCL kernel for each layer of a network. 
For large networks, this approach results in a non-fitting designs, 
mostly due to the resources consumed by the LSUs, especially 
when the kernel has been tiled and unrolled.

Thus, we group and parameterize similar kernels so that their
hardware may be reused across layers of the network. 
Specifically, we group operations by the filter size and stride 
of convolutions. The number of filters, input channels, and the 
input feature map spatial dimensions are the parameters (kernel 
arguments) and can be set at runtime to execute different layers 
of the network. This optimization is implemented using symbolic 
shape execution in TVM.

\subsection{Optimized Float Operations}
\label{sec:float}

We utilize two AOC flags that improve the generation for floating
point operations.  The first is \texttt{-fp-relaxed}, which
relaxes the order of floating point operations, allowing them to
be more efficiently implemented in hardware.
The second is the \texttt{-fpc} flag that removes intermediate
rounding operations and conversions, and only performs them once
at the end. This may result in fused floating-point operations,
such as fused multiply-accumulate (FMAC), which potentially
reduces area usage.

\subsection{Optimizations Application}
\label{sec:application}

\begin{table*}
\centering
\footnotesize
\setlength{\tabcolsep}{0pt}
\begin{tabular*}{1\textwidth}{@{\extracolsep{\fill}}llllll@{}}
\hline
Optimization & Pipelined & Folded & Parameters & Pattern \\ \hline
Loop unrolling (LU) & \checkmark & \checkmark & Unroll factor & All kernels except transpose/padding \\
Loop fusion (LF) & \checkmark & \checkmark & & Activation/batchnorm in Conv, FC, pooling \\
Cached Writes (CW) & \checkmark & \checkmark & & All kernels except transpose/padding \\
Optimized Float Operations (OF) & \checkmark & \checkmark & & \texttt{-fpc -fp-relaxed} for all bitstreams \\ 
Channelization (CH) & \checkmark & & Channel depth & Movement of activations, all layers \\
Autorun Kernels (AR) & \checkmark & & & Pooling, transpose/padding \\
Concurrent Execution (CE) & \checkmark & & & Host optimization \\
Parameterized Kernels (PK) & & \checkmark & & Convs with same stride and filter size \\
Loop tiling (LT) & & \checkmark & Tile factor & Conv, FC \\
\end{tabular*}
\caption{Summary of optimizations and their applicability} 
\label{tbl:optimizations:summary}
\end{table*}

We employ a pattern-based approach to applying the optimizations.
Table~\ref{tbl:optimizations:summary} lists the optimizations, the
execution mode in which they are utilized (i.e., pipelined or 
folded), the parameters of the optimization (where present), and
the pattern to which an optimization is applied. For example, we 
fuse the loops for activations and batch normalizations to the 
convolution loops when the opportunity is present. 
Loop unrolling, fusion, 
and float optimizations are 
applied in both pipelined and folded execution modes. However, the 
application of the other optimizations depends on the mode of execution. 

Channels are used to implement kernel-to-kernel transfer of output 
feature maps in pipelined execution. Similarly, kernels that represent 
layers without weights or biases are declared autorun. Further, separate 
command queues can be created for each kernel in pipelined kernels to 
implement concurrent execution. 

In parameterized execution, these optimizations are not applicable. 
Since kernels can be used more than once, there is no longer a 
single downstream path for output feature maps and must be stored 
in global memory. Thus, channels are not used, and because kernels 
must read feature maps from global memory, they cannot be declared 
autorun. Further, since channels were used to synchronize kernels 
executing concurrently, concurrent execution is also not enabled 
for parameterized kernels either. Instead, we opt to tile and unroll 
in multiple dimensions.

The application of the optimizations require the specification of 
three factors: unrolling factors, tiling factors for parameterized 
kernels, and channel depth for pipelined kernels. 
Since we unroll all inner loops resulting from strip 
mining/tiling, the unrolling and tiling factors are 
the same in this context. We set three requirements for 
choosing a factor:

\begin{enumerate}
\item For loops that access global memory and the data is not 
cached, the unrolling factor should not exceed theoretical 
peak external memory bandwidth of the FPGA. For example, the 
Stratix~10SX has a theoretical bandwidth of 76.8 GB/s. Assuming 
a 250~MHz operating frequency, this can support 307.2 bytes/cycle, 
which is approximately 76 floats. Therefore, the factor should 
not exceed 76 on this target platform\footnote{Observed by others 
in previous work as the bandwidth roof, part of the roofline 
model~\cite{Zhang:2015:OFA:2684746.2689060}.}. 
\item The loop count must be evenly divisible by the factor to
avoid prologues and epilogues.
\item The design must not exceed device resources.
\end{enumerate}

The estimation of the resources of a design can be time consuming.
While DSP usage can be predicted by counting the number of 
multiply-accumulate operations and multiplying by the unroll/tile 
factors, it is not straight forward to predict logic and BRAM usage 
based on the OpenCL code. Further, AOC often grossly overestimates 
logic usage. Thus, it becomes necessary to utilize Quartus' place
and route to get more accurate resource usage, including that for
routing. This is time consuming, and can take from 3 to 12 hours, 
depending on the size of the FPGA and the complexity of the design. 
Thus, we manually sweep through several parameter values and select
ones that adhere to the above requirements. Ideally, a design space
explorer (DSE) can be developed to automate this process or possible
augment it with a model-based prediction of resources. We leave 
such DSE to future work.

Finally, a buffer depth must also be specified for buffered channels. 
The depth must be sufficient to hold the output of the largest feature
map in the network. For example, in LeNet-5, the channel depths 
must be large enough to fit the output feature map of a convolution
layer, which is 256 floats, or 1024 bytes.

%% file: evaluation.tex
\section{Evaluation}
\label{sec:evaluation}
 
We experimentally evaluate the impact of the optimizations and 
the performance of the accelerators generated by our flow. We place 
this performance in context by comparing it to that of a CPU 
and a GPU. Finally, we compare to the performance of related 
approaches to expose the strengths and weaknesses of our approach.

\subsection{Networks}
\label{sec:networks}

We use three networks: LeNet-5, MobileNetV1 and ResNet-34. We 
define LeNet-5 in Keras and train it with the MNIST data set.
We use the MobileNetV1 network 
architecture and pretrained ImageNet parameters from Keras 
Applications~\cite{KerasApplications}, a library of DNN models 
and pretrained weights. Lastly, for ResNet-34 we use the models 
and pretrained parameters from the \texttt{image-classifiers} 
Python library.  All networks use 32-bit floating-point arithmetic.

\subsection{Platforms}
\label{sec:platforms}

We conduct our experiments on a PCIe D5005 Programmable 
Acceleration Card (PAC) with a Stratix~10SX FPGA 
(1SX280HN2F43E2VG), provided by the Intel Labs Academic 
Compute Environment~\cite{ilace}. This S10SX device has
over 1.6M ALUTs, 3.4M FFs, 5.7K DSPs and 11M bits
of on-chip RAM. It also has 32GB of external DDR4 arranged 
in 4 banks, with a theoretical peak bandwidth of 76.8GB/s. 
It is hosted by a 28-core Intel Xeon Platinum 8180 processor 
with 384GB of memory and a Gen~3 PCIe.

We evaluate CPU performance on server-grade Intel Xeon 
Platinum 8280s with two physical CPUs, for a total of 56 
cores and 112 threads and 768 GB DDR4 memory. We use an 
NVIDIA GTX~1060 with 6 GB of VRAM for the 
GPU evaluations. The CPU performance is measured using the 
LLVM-CPU backend in TVM executing with $n$ threads, labeled 
as TVM-$n$t. It is also measured using TensorFlow (labeled TF).
The number of threads TensorFlow uses is automatically 
determined by the framework to optimize performance. Thus,
it can vary up the the maximum number of threads. The GPU 
performance is measured using TensorFlow 
for the GPU, with the CUDA Deep Neural Network Library (cuDNN) 
and is labeled TF-cuDNN. We use TVM release v0.7 (commit 728b829), 
Keras 2.3.1, and TensorFlow 2.1.0.

\subsection{Metrics}
\label{sec:metrics}

The key metric we use for our evaluation of inference performance
is {\em Frames per second} (FPS), defined as the number of forward 
passes that can be processed in a second. It is obtained by measuring 
the execution time $t$ it takes to classify $N$ 
images\footnote{We use N=1000 images.} and then calculating $N/t$.
The time measurement is made using the OpenCL kernel event profiler. 
When comparing to existing work, we calculate floating point
performance in GFLOPS, which is computed using FPS and the number 
of floating point operations performed by the networks we accelerate.

\subsection{Results}
\label{sec:results}

Table~\ref{tbl:resource} lists the $f_{max}$ and resources used by 
the accelerators generated by our flow for each of the three networks.
In all three cases, we observe that the generated circuits utilize 
a fraction of the resources, particularly the DSPs. This underutilization
of DSPs limits performance as discussed in Section~\ref{sec:discussion}
below.

\begin{table}[h!]
\centering
\begin{tabular}{|c|c|c|c|c|}
\hline
           & Logic (\%)  & BRAM (\%) & DSP (\%) & $f_{max}$ \\ \hline
LeNet-5    & 25\%        & 19\%      & 5\%      & 218  \\ \hline 
MobileNetV1& 46\%        & 48\%      & 15\%     & 187  \\ \hline
ResNet-34  & 59\%        & 61\%      & 16\%     & 125  \\ \hline
\end{tabular}
\caption{Resource utilization and $f_{max}$ (MHz)} 
\label{tbl:resource}
\end{table}

Table~\ref{tbl:optimizations} shows the optimizations applied to
each network (see Table~\ref{tbl:optimizations:summary} for
optimization name abbreviations). In the case of LeNet-5, the network is small and
thus can fit in its entirety on the S10SX. Thus parameterized
kernels (PK) are not used. However, they are used for the larger
MobileNetV1 and ResNet-34, precluding the use of CH, AR and CE, as
was explained in Section~\ref{sec:optimizations}.

\begin{table}[h!]
\addtolength{\tabcolsep}{-1pt}
\centering
\begin{tabular}{|c|c|c|c|c|c|c|c|c|c|c|}
\hline
           & PK         & LU         & LT         & LF         & CW         & OF         & CH         & AR         & CE          \\ \hline
LeNet-5    &            & \checkmark &            & \checkmark & \checkmark & \checkmark & \checkmark & \checkmark & \checkmark  \\ \hline
MobileNetV1& \checkmark & \checkmark & \checkmark & \checkmark & \checkmark & \checkmark &            &            &             \\ \hline
ResNet-34  & \checkmark & \checkmark & \checkmark & \checkmark & \checkmark & \checkmark &            &            &             \\ \hline
\end{tabular}
\caption{Applied Optimizations}
\label{tbl:optimizations}
\end{table}

Table~\ref{tbl:basevsopt} reports the FPS obtained for both
the base (unoptimized) accelerators and the optimized ones.
It also shows the speedups achieved using the 
optimizations. The table shows that the performance of the
base accelerators is poor and the optimizations deliver
significant speedups, up to 846$\times$ for ResNet-34. The 
demonstrates the effectiveness of the optimizations in
improving performance.

\begin{table}[h!]
\centering
\begin{tabular}{|c|c|c|c|}
\hline
           & Base     & Optimized & Speedup  \\ \hline
LeNet-5    &  524     & 4917      & 9.38$\times$    \\ \hline
MobileNetV1&  0.17    & 30.3      & 178.2$\times$   \\ \hline
ResNet-34  &  8.3e-3  & 7.04      & 846$\times$     \\ \hline
\end{tabular}
\caption{FPS of base versus optimized circuits}
\label{tbl:basevsopt}
\end{table}

\begin{table*}[th]
\centering
\begin{tabular}{|c|c|c|c|c|c|}
\hline
           & S10SX  & \multicolumn{3}{c|}{CPU}    & GPU           \\ \hline
           &        & TVM-1t              & TVM-56t             & TF                       &  TF-cuDNN     \\ \hline
LeNet-5    &  4917  & 2345 (2.10$\times$) & 1470 (3.34$\times$) & 1075 (4.57$\times$)& 1604 (3.07$\times$) \\ \hline
MobileNetV1&  30.3  & 15.6 (1.94$\times$) & 84.5 (0.36$\times$) & 21.6 (1.4$\times$) & 43.7 (0.69$\times$) \\ \hline
ResNet-34  &  4.6   &  1.2 (3.83$\times$) & 13.7 (0.34$\times$) & 10.7 (0.43$\times$)& 31.7 (0.15$\times$) \\ \hline
\end{tabular}
\caption{FPS (Speedup) comparisons to CPU and GPU}
\label{tbl:compare}
\end{table*}

Table~\ref{tbl:compare} gives the performance and the speedups
of our accelerators over those of the CPU and the GPU. Compared
to TVM, the accelerators outperform the single-threaded TVM
by up to 3.83$\times$. When the number of threads is increased,
our accelerators outperform TVM for LeNet-5 but underperform
for MobileNetV1 and ResNet-34. This under-performance is 
expected given the large number of threads. Similarly,
for TensorFlow, our accelerators outperform the framework
for LeNet-5 and MobileNetV1 but underperform for ResNet-34,
presumably because TensorFlow is using more threads for this
network. Both TensorFlow and TVM are highly optimized ML frameworks 
and are considered state-of-the-art CPU implementations. The
speedups demonstrate that our optimizations can make the
performance of our automatically generated FPGA kernels 
surpass that of optimized CPU implementations.

Our accelerators outperform the GTX~1060 for LeNet-5, even though
the GPU implementation uses the highly optimized cuDNN framework. 
This is likely because the network is small and has few parameters. 
The larger memory bandwidth on GPUs does not provide an advantage since 
the weights can be stored in on-chip caches with the FPGAs. Further, 
since batching is not used, it is possible that the GPU is underutilized 
for a network of this size whereas we are able to have increased 
utilization with layer-pipelined execution (i.e, channels and 
concurrent execution). Therefore, the FPGA can demonstrate superior 
performance. However, for the larger networks, the larger memory 
bandwidth of the GPU makes its performance superior to that of our 
accelerators, particularly for ResNet-34. 

\subsection{Comparison to Existing Work}

We compare the performance of our accelerators to those of three
related works: Caffeinated FPGAs (DiCecco et al.~\cite{7929549}), 
TensorFlow to Cloud FPGAs (Hadjis et al.~\cite{8892010}), and 
DNNWeaver (Sharma et al.~\cite{DNNWeaver}). Caffeinated FPGAs is 
a modification of the Caffe ML framework with support for an
HLS-generated hand-optimized FPGA Winograd $3\times 3$ convolution 
engine in OpenCL. TensorFlow to Cloud FPGAs is a closely related 
framework that allows the compilation of TensorFlow models to 
Amazon FPGA devices using a hardware IR called Spatial. DNNWeaver 
is an accelerator generation framework constructed from 
hand-optimized hardware templates 
implemented in RTL with a design space explorer. We compare to these 
works to determine the competitiveness of our compiler flow against 
respectively a hand-optimized HLS approach, a hardware IR generation 
approach (a higher abstraction than OpenCL HLS), and an RTL 
generation approach from hand-optimized parameterized RTL components.

We report 70.4 GFLOPS for our 3$\times$3 convolutions
in ResNet-34, which compares to a geometric mean of the 50
GFLOPS reported by DiCecco et al.~\cite{7929549}, for a speedup of
1.4$\times$.  However, considering the technology gap with the previous 
generation FPGA used in their experiments, we consider this 
improvement marginal. Nonetheless, our flow can support any 
size convolution with no hand-optimizations, while theirs supports 
only 3$\times$3 filter sizes and is hand-optimized.

Hadjis et al..~\cite{8892010} benchmarks LeNet-5 on 
the Xilinx UltraScale+ VU9P FPGA. 
They report 3.49~GFLOPS, but assume 2.29M FP operations. In 
contrast, we calculate only 389K FP operations. This suggests that 
their implementation and ours is somewhat different. Normalizing 
to the number of FP operations, they achieve 0.59~GFLOPS compared 
to our 1.91~GFLOPS, which is $3.23\times$ faster.

Finally, Venieris et al.~\cite{10.1145/3186332} include DNNWeaver in 
their survey. We compare their AlexNet (1.33G FP operations) performance 
with our MobileNetV1 (1.11G FP operations). Their AlexNet accelerator is
faster by $9.22\times$. This is not surprising since they utilize
highly-optimized RTL templates and employ a design space explorer.

\subsection{Discussion}
\label{sec:discussion}

Our results show that our optimizations are effective in improving 
the performance of the generated hardware. Indeed, this performance
is better than or competitive with that of TensorFlow and TVM, both
highly optimized CPU frameworks. Only for larger networks and with
the use of many threads, do our accelerators underperform
the frameworks. While our accelerators exhibit better performance
compared to other frameworks that use HLS, it underperforms hand-designed
and optimized designs.
 
A key factor that limits the performance of our accelerators is the 
underutilization of the DSPs, particularly for larger networks. This 
underutilization reflects that more computations can performed 
in parallel, thus improving performance. Higher use of DSPs can be 
achieved with larger tiling factors, which allow for more unrolling.

However, increasing the tiling factors gives rise to other bottlenecks
that act to degrade performance. Routing congestion increases with larger 
tile sizes, leading to large drops in $f_{max}$. The congestion can also lead
to routing failure before utilizing all DSPs. Specifically, 
larger tile sizes lead to either more LSUs or wider LSUs created for feeding
the DSPs with weights and feature maps. The fanout from these LSUs can 
lead to the routing failure. This is the case for both MobileNetV1 and 
ResNet-34. Similarly, the increased number of LSUs leads to more BRAM and 
logic overhead that prevents scaling the designs with more DSPs. This is 
the case for ResNet-34.

The above bottlenecks can be mitigated by using vector types to align loads/stores,
which will reduce the logic and memory overhead incurred from LSUs, exploring deployments
that use a mix of pipelined and folded execution, and using reduced precision arithmetic to 
fit more operations per DSP and alleviate memory requirements. We leave addressing these 
limitations to future work.

%% file: related.tex
\section{Related Work}
\label{sec:related}

There is a large volume of work on FPGA acceleration of DNNs\footnote{
Over 330 published works in major FPGA conferences.}. We focus on
work that is most related to ours, namely automated design flows.
Early work focused on accelerators that are fixed for a specific workload 
(such as a model or an operation) or a specific FPGA 
target~\cite{Zhang:2015:OFA:2684746.2689060, Qiao2017, 7929549}.

More recent work added flexibility to map CNNs from a high-level description 
to either a fixed architecture or one that is dynamically generated 
for a specific model. Zhang et al.~\cite{Zhang:2016:CTU:2966986.2967011} 
present Caffeine, an accelerator design guided by the roofline model. 
It's flow allows for software-definable parameters (e.g., convolution filter 
size) and hardware-definable parameters (e.g., BRAM size and kernel size). 
Qiao et al.~\cite{Qiao2017} similarly design an accelerator with a matrix 
multiplication engine designed for the Xilinx Zynq where convolutions 
are mapped to the matrix multiply PEs.

DeepBurning~\cite{Wang:2016:DAG:2897937.2898003} is a Caffe-oriented 
design flow that maps CNN layers to building blocks in a CNN components
library, written in RTL. The work employs both spatial and temporal
folding to map layers to the blocks. 

DNNWeaver~\cite{DNNWeaver} is an automated design flow that generates 
an accelerator from a Caffe model description with hand-optimized 
template designs. The architecture includes Processing Units (PUs), 
made up of smaller PEs that implement convolutions and inner products, 
with other custom modules that execute pooling and activation layers. 

Venieris et al.~\cite{venieris2016fccm} describe fpgaConvNet, a 
flow that models CNNs as synchronous dataflows (SDF). They also define
transformations to partition and coarse-grain fold operations which 
effectively unrolls independent operations. In addition, they perform
fine-grain folding, which involves unrolling reductions, time-multiplexing 
a single MACC unit filter width $\times$ height times to compute dot 
products in one cycle. 

Intel DLA~\cite{DBLP:conf/fpl/AbdelfattahHBDO18} is an FPGA overlay 
written in OpenCL as a dot product engines mapped to by a proprietary 
graph compiler. It employs optimizations to increase utilization and deliver
state-of-the-art performance. However, new operators require the addition 
of new functional units to the overlay and it is unclear how to 
provide such units, limiting flexibility. 

Hadjis et al.~\cite{8892010} use an HDL abstraction to deploy on both 
Xilinx and Intel FPGAs, providing more portability than using Intel 
tools. They allocate DSPs proportionally to the number of 
multiply-accumulate operations associated with a layer.

HPIPE~\cite{10.1145/3373087.3375380} is a FPGA sparse network acceleration 
architecture and graph compiler. Parameters for parallelism are selected 
algorithmically to balances the throughputs of all layers while 
maximizing DSP utilization. HPIPE is capable of zero-skipping but stashes 
all weights on-chip, limiting it to models for which parameters that can 
fit into on-chip memory. They prune the network to mitigate this limitation.

Our approach differs from the above work in several ways.
The above work relies on hardware templates, either in RTL or for HLS. 
This requires these templates to be extended for new model operations;
a time-consuming process that involves hardware design and optimization.
In contrast, we directly generate hardware for a given network. Adding 
a new operation in our approach involves a high-level description in 
TVM compute functions and schedules. Second, much of the above work 
focuses on specific ML frameworks (Caffe for the most part) while 
we use TVM to take advantage of its support for a variety of 
frontends. Lastly the above work performs more aggressive optimizations 
for their templates. The use of optimized templates with these 
optimizations leads to higher performance. Our evaluation provides 
a comparison to representatives of this work.

Our work extends our earlier exploratory study on the approach~\cite{raw20}.
It extends this earlier work by the expanding the optimizations,
automating them in TVM and by evaluating the impact of these
automated optimizations on performance.

There is also work on optimizing OpenCL programs for execution on
FPGAs. For example, Sanaullah et al.~\cite{Sanaullah2018} present 
empirically guided code optimizations for hand-written OpenCL programs 
and show their impact on performance. Our optimizations are not
unlike the ones they describe, but we focus on compiler-generated
kernels and automate their application. In either case, the application
of the optimizations is guided by the best practices provided by the 
AOC compiler~\cite{IntelProgrammingManual}.

%% file: conc.tex
\section{Concluding Remarks}
\label{sec:conc}

We propose, implement and evaluate a compiler-based flow for generating 
CNN inference accelerators on FPGAs. The flow uses TVM to generate OpenCL 
kernels, which are then optimized for parallelism, memory utilization 
and resource usage. The kernels are converted into RTL using Intel's AOC 
compiler and synthesized for the target FPGA. The flow offers several 
benefits, including the use of industry standard tools, generating hardware 
without the need to maintain hardware components libraries and portability 
across FPGA targets. 

We describe the optimizations we automate within TVM, which include
unrolling, tiling, and fusion. We also utilize kernel 
parameterization to allow for the use of the same kernel hardware across 
layers, thus generated designs to fit within available resources. Finally, 
we apply a number of modifications to the host program to allow for 
channelization, auto-run kernels and concurrent execution.

We evaluate the impact of the optimizations and the performance our 
accelerators. We show that the optimization can improve the 
performance of the generated hardware by up to 846$\times$ over the base 
kernels generated by TVM. The performance of the accelerators is competitive 
with CPU TensorFlow and TVM, and with some existing hand-optimized or 
template-based approaches. Although in other cases the accelerators do not 
outperform hand-optimized designs, we view our flow as useful in pre-production 
environments that benefit from fast prototyping and increased performance.

There are several directions in which this work can be extended.
First, quantized networks that reducing bit precision for weight/activation 
representation can be supported. Second, the work can be extended to support 
sparse computations. Third, a design space exploration framework can aid
in the application of the optimizations and in assessing the potential 
trade-offs that result from their collective application. Finally, support
for multi-FPGA devices can aid in generating accelerators for larger 
networks.